\documentclass[floatfix, prb, preprint,showpacs]{revtex4}

\usepackage{graphicx}

\usepackage{dcolumn}
\usepackage{longtable}
\usepackage{bm}

\def\LCO/{LiCu$_2$O$_2$}
\def\wn/{\,cm$^{-1}$}
\def\DM/{Dzyaloshinskii-Moriya}

\begin{document}

\preprint{\today}
\title{Magnetic excitations and optical transitions in the multiferroic spin-$\frac{1} {2}$ system LiCu$_2$O$_2$
}
\author{D.~H{\"u}vonen}
\author{U.~Nagel}
\author{T.~R{\~o}{\~o}m}

\affiliation{National Institute of Chemical Physics and Biophysics,
Akadeemia tee 23, 12618 Tallinn, Estonia.}


\author{Y.J.~Choi}
\author{C.L.~Zhang}
\author{S.~Park}
\author{S.-W.~Cheong}
\affiliation{Rutgers Center for Emergent Materials \& Department of Physics and Astronomy, Rutgers University,
Piscataway, New Jersey 08854, USA}

\date{\today }

\begin{abstract}
Magnetic excitations in cycloidal magnet LiCu$_2$O$_2$  are explored using THz absorption spectroscopy in magnetic fields up to 12\,T.
Below the spin ordering temperature we observe eight optically active transitions in the spin system of LiCu$_2$O$_2$ in the energy range from 4 to 30\,cm$^{-1}$. 
In magnetic field the  number of modes increases and the electric polarization flop is seen as a change in magnetic field dependence of mode energies.
The polarization dependence of two of the modes in zero magnetic field fits the selection rules for the cycloid tilted by $\theta=41\pm1^{\circ}$ from the bc plane.
For the remaining six modes electric and magnetic dipole approximations cannot explain the observed polarization dependence.
We do not see the electromagnon in the explored energy range although there is evidence that it could exist below 4\,cm$^{-1}$.

\end{abstract}

\pacs{75.30.Ds, 76.50.+g, 75.25.+z, 75.50.-y, 78.30.-j, 78.20.Ci}

\maketitle

The coupling between magnetic and electric orders in multiferroic materials makes them attractive for technological applications.
An especially interesting   class of multiferroics (MF) are frustrated magnets, where the charge order is driven by the magnetic order \cite{Cheong2007} and even moderate magnetic fields are sufficient to select between different magnetic ground states and change the charge order due to magnetoelectric (ME) coupling.
The ME interaction transfers part of  the electric dipole moment of the charge-ordered state to the spin waves enhancing their ability  to absorb electromagnetic waves, typically  in the THz range.\cite{Katsura2007,Aguilar2009}
A priori, the knowledge of the strength of the static ME coupling for a given material does not tell us the strength of the  dynamic ME coupling.
Therefore the exploration of THz properties is of interest to fully exploit MF materials.

\begin{figure}[tb]
\includegraphics[width=7.6cm]{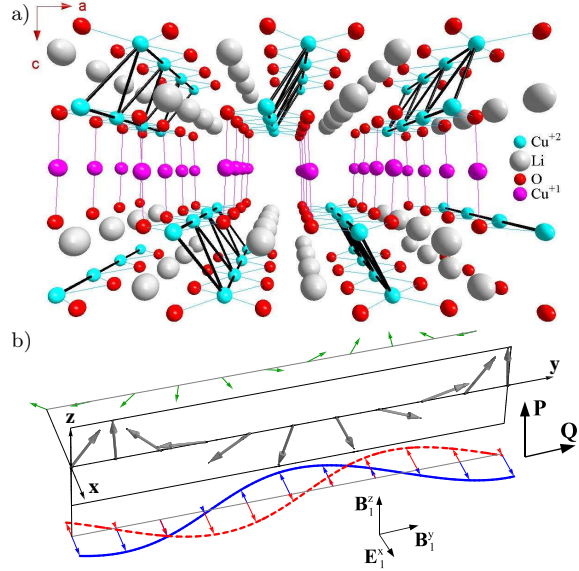}
\caption{(color online) 
(a) Crystal structure of \LCO/. Cu$^{2+} $ spins, joined by thick black lines,  form a zig-zag spin ladder.
(b) Cycloidal order of spins $ \mathbf{S}_i $ positioned along the $y$ axis, shown with thick solid arrows.  
Small arrows show the change  $ \delta \mathbf{S}_i $ to visualize the spin wave dynamics.
The $\omega_y$ and $\omega_z$ modes change $ \mathbf{S}_i $ in $x$ direction and are drawn by blue arrows (solid  line envelope) and red arrows (dashed  line envelope), respectively.
Green arrows in the $yz$ plane show the $\omega_x$ mode.
}
\label{fig1}
\end{figure}

The spin cycloid is an example of a frustrated magnetic structure where ME coupling is allowed.
In the spin cycloid the electric polarization is induced either through the spin current \cite{Katsura2005} or the inverse \DM/ (DM) interaction mechanism.\cite{Sergienko2006,Mostovoy2006}
In an one-dimensional chain of spins along the $y$ axis with spins rotating in the $yz$ plane the electric polarization $\mathbf{P}=(0,0,P)$ is perpendicular to the cycloidal ordering vector $\mathbf{Q}=(0,Q,0)$ and perpendicular to the normal of the spin rotation plane, $\mathbf{P} \propto \mathbf{Q} \times (\mathbf{S}_i \times \mathbf{S}_{i+1})$, see Fig.\,\ref{fig1}b.
The orientation of the spin rotation plane is determined by anisotropic interactions, like easy plane anisotropy  or DM interaction.
Several multiferroic materials have a cycloidal spin order: 
 $\mathrm{TbMnO_3}$,  $\mathrm{DyMnO_3}$ [\onlinecite{Kimura2003,Goto2004,Kimura2005}],  $\mathrm{Ni_3V_2O_8}$ [\onlinecite{Lawes2005}], \LCO/ [\onlinecite{Park2007}],  $\mathrm{LiVCuO_4}$ [\onlinecite{Naito2007}], and  CuO [\onlinecite{Kimura2008}]. 
Among them, \LCO/ was the first low spin, $S=1/2$, system to show electric polarization in the cycloidal phase.
Although \LCO/ seems to be a good model system with  prevailing one-dimensionality of its magnetism, the connection between the electric polarization and the underlying magnetic structure is not well understood. 

The structure of \LCO/, shown in Fig.\,\ref{fig1}a,  contains Cu-O chains extending along the crystallographic $b$ axis. 
The chains form a zig-zag Cu$^{2+}$ $S=1/2$ spin ladder with the ladder plane tilted out from the $bc$ plane.\cite{Berger1992}
The ladders in the planes, stacked along the $c$ axis, have an alternating tilt angle $\pm 38.4^{\circ}$ leading to two inequivalent ladders per unit cell.
The first magnetic phase transition is at 24.6\,K into the sinusoidal spin structure.\cite{Rusydi2008}
An incommensurate cycloidal spin order with $\mathbf Q \!\parallel\! \mathbf b$, driven by frustrated isotropic exchange interactions inside the  ladders, appears in the second phase transition\cite{Rusydi2008,Masuda2004,Gippius2004} below 23\,K together with  spontaneous polarization\cite{Park2007} $\mathbf{P} \!\parallel\!  \mathbf{c}$.
The knowledge about the orientation of the cycloid plane is controversial.   
Initially an $ab$ plane cycloid was considered in the neutron scattering study.\cite{Masuda2004,Masuda2005}
However, the direction of $\mathbf{P}$  together with the  inverse DM mechanism dictates a $bc$ plane cycloid.\cite{Park2007}
A later study by Seki \textit{et al}. \cite{Seki2008} confirmed the  $bc$ plane cycloid and $\mathbf{P}\!\parallel\! \mathbf{c}$.
However, they found deviations from the $bc$ cycloid in accord with the NMR study.\cite{Gippius2004}
The presence of ME coupling in \LCO/ is demonstrated by the flop of polarization  from $\mathbf{P}\!\parallel\! \mathbf{c}$  to $\mathbf{P}\!\parallel\! \mathbf{a}$ if magnetic field $\mathbf{B}_0>2$\,T is applied along $b$ axis.\cite{Park2007}
Such a flop can be explained by invoking a ME coupling between two spin chains in neighboring ladder planes.\cite{Fang2009}
Also, resonant soft x-ray magnetic scattering shows evidence that the inter-chain spin coupling  along the $c$ direction  is essential for  inducing polarization.\cite{Huang2008}

Modes at wavevectors $\mathbf{k}=0$  and $\mathbf{k}=\pm \mathbf{Q}$ couple to long-wavelength electromagnetic radiation within the linear spin wave model of a spin cycloid.
They correspond,  as shown in Fig.\,\ref{fig1}b,  to the  rotations of spins about three  axes by small angles\cite{Katsura2007}.
The rotation about the $x$ axis ($\mathbf{k}=0$) costs no energy because the spins rotate within the easy plane, hence $\omega_x=0$.
Rotations about two other axes  have finite energies, $\omega_y=\omega_z$,  the magnitude determined by the size of  the easy plane anisotropy.
The motion of spins induced by magnetic component $\mathbf{B}_1$ of THz radiation is described by
$ \delta \mathbf{S}_i /  \delta t  \propto \mathbf{B}_1 \times  \mathbf{S}_i $.
Accordingly, all three modes are magnetic dipole active, 
$\mathbf{B}_1^{y}$ excites $\omega_y$ and $\mathbf{B}_1^{z}$ excites $\omega_z$ with equal intensity in zero field;
$\mathbf{B}_1^{x}$ excites $\omega_x$ although this mode has zero energy.
In the presence of dynamic ME coupling the $\omega_y$ mode is excited by the electric component $\mathbf{E}_1^{x}$ of THz radiation\cite{Katsura2007} and is referred to as the electromagnon.\cite{Pimenov2006Nature}
Thus, the orientation of the spin cycloid can be determined from the polarization dependence of THz absorption.

In this Communication  we study magnetic field and polarization dependence of the THz absorption in the cycloidal phase of \LCO/.
Our aim is to find the electromagnon, the orientation of the spin cycloid, and changes in the spin excitation spectrum upon the polarization flop in 2\,T field. 
Previously only one THz absorption study on \LCO/  has been reported\cite{Mihaly2006}, the experiment being limited to $\mathbf{B}_0\!\parallel\! \mathbf{c}$, where a peak at 12\wn/ with increasing energy in the field was found.

We use three single crystal samples  from the same batch as  in Ref.\,\onlinecite{Park2007}.
Sample L1 has $ab$ plane area of 6.16\,mm$^{2}$ and thickness of 0.75\,mm in $c$ direction, L2 and L3 have $ac$ plane areas 1.44\,mm$^{2}$ and  6.88\,mm$^{2}$ with thicknesses 1.75\,mm and  4.3\,mm in $b$ direction, respectively. 
Since the crystals are twinned in the $ab$ plane, with a twin size of the order of 10\,$\mu$m, we use $a(b)$ notation when referring to direction-sensitive physical properties. 
THz absorption spectra are recorded using a Martin-Puplett spectrometer with a 0.3\,K bolometer and a rotatable polarizer in front of the sample.\cite{room2004NaVa}

\begin{figure}[tb]
\includegraphics[width=8.3cm]{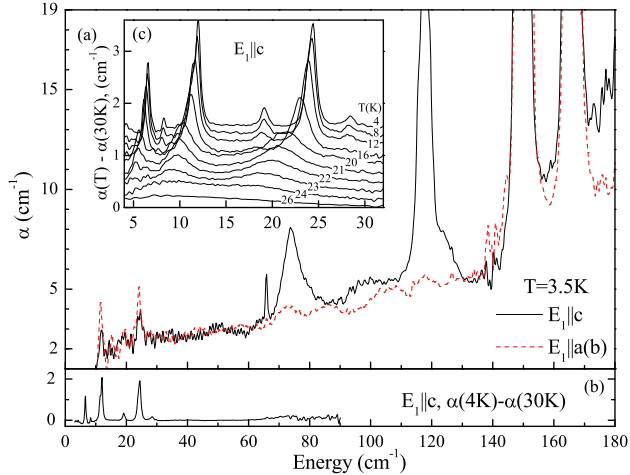}
\caption{(color online) Light absorption in  LiCu$_2$O$_2$ (sample L3) below 180\,cm$^{-1}$  measured  in $\mathbf{E}_{1}\!\parallel\!\mathbf{a(b)}$  (dashed red line) and   $\mathbf{E}_{1}\!\parallel\!\mathbf{c}$ (solid black line) polarizations. $T$ dependence of absorption, $\mathbf{E}_{1}\!\parallel\!\mathbf{c}$, is shown by differential absorption spectra in (b) and in  the inset (c) .}
\label{LiAbsTdep}
\end{figure}

\begin{figure}[tb]
\includegraphics[width=8cm]{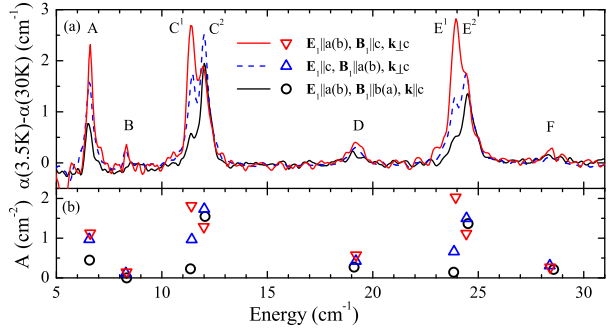}
\caption{(color online) Polarization dependence of  absorption in LiCu$_2$O$_2$ below 35\,cm$^{-1}$, $T=3.5$\,K, $B_{0}=0 $\,T.
(a) Differential absorption spectra are shown by curves  and (b) absorption line areas from Lorentzian fits are given by symbols  for following polarizations:
black  line and circles for $\mathbf{E}_{1}\!\parallel\!\mathbf{a(b)}, \mathbf{B}_{1}\!\parallel\!\mathbf{b(a)}$ is an average of spectra measured on samples L1 and L2;
red  line and down triangles for $\mathbf{E}_{1}\!\parallel\!\mathbf{a(b)}, \mathbf{B}_{1}\!\parallel\!\mathbf{c}$, sample L2; 
blue dashed line and up triangles for $\mathbf{E}_{1}\!\parallel\!\mathbf{c}, \mathbf{B}_{1}\!\parallel\!\mathbf{a(b)}$, sample L2.
The letters from A to F in (a) label lines, see Table\,\ref{LiTable}.
}
\label{LiAbsPolDep}
\end{figure}

The absolute spectra in Fig.\,\ref{LiAbsTdep}a show several strong phonon absorption lines. 
The differential spectrum in Fig.\,\ref{LiAbsTdep}b shows that only the lines  below 40 \wn/ that disappear above the cycloidal ordering temperature $T=23$\,K are spin wave excitations.
In the rest of the paper we use the 0\,T  spectrum at 30\,K as a reference spectrum by subtracting it from the spectra measured in the given field and at lower $T$. 

\begin{figure}[tb]
\includegraphics[width=8cm]{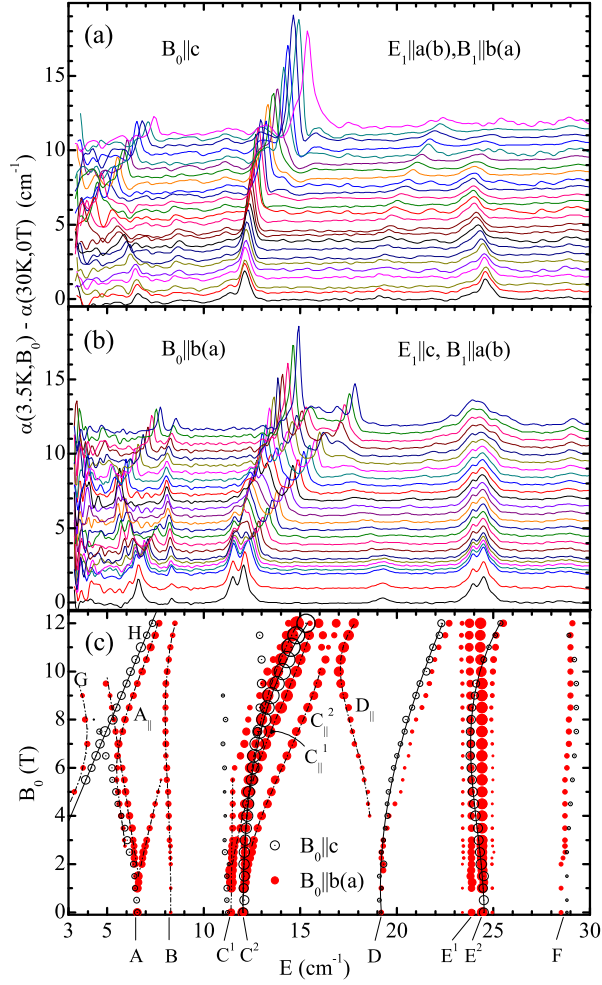}
\caption{(color online) $\mathbf{B}_0$ dependence of THz absorption in  LiCu$_2$O$_2$ at 3.5\,K.
Differential absorption spectra (a)  $\mathbf{B}_0 \!\parallel\! \mathbf{c}$ and $\mathbf{E}_1 \!\parallel\! \mathbf{a(b)}$, $\mathbf{B}_1 \!\parallel\! \mathbf{b(a)}$, sample L1 ; (b)  $\mathbf{B}_0 \!\parallel\! \mathbf{b(a)}$,  $\mathbf{E}_1 \!\parallel\! \mathbf{c}$, $\mathbf{B}_1 \!\parallel\! \mathbf{a(b)}$, sample L2. Panel (c) has the results of Lorentzian fits where the area of the symbol is proportional to the absorption line area $A$; black empty circles $\mathbf{B}_0  \!\parallel\! \mathbf{c}$ and  filled red circles $\mathbf{B}_0 \!\parallel\! \mathbf{b(a)}$. Black solid and dash-dot lines are eye guides, except line H which is a linear fit of line position as a function of $B_0$.}
\label{BdepLinePos}
\end{figure}

Two peaks are emerging already in the 24\,K spectrum in the sinusoidal phase and are easily seen at 23\,K (Fig.\,\ref{LiAbsTdep}c).
As $T$ drops the peak energies go up and eventually at least six peaks can be seen in the 4\,K spectrum.
Higher resolution spectra show that the two strongest peaks at 12 and 24\wn/ are doublets, split by  0.8\wn/, Fig.\,\ref{LiAbsPolDep}.
This increases the total number of peaks to eight (Table\,\ref{LiTable}).
As $\mathbf{B}_0$  is applied, Fig.\,\ref{BdepLinePos}, lines shift,  change their intensity, and some split.
In fields stronger than 5\,T  new modes, G and H, emerge into the spectral window from the low energy side, Fig.\,\ref{BdepLinePos}c.
The expected number of optically active modes within the spin wave approximation is two, $\omega_y$ and $\omega_z$.
There are two spin chains in the unit cell and the coupling between the chains in $c$ direction may split C and E modes into C$^1$, C$^{2}$ and E$^{1}$, E$^{2}$.
In addition, if the cycloid is not circular, ellipticity activates otherwise optically silent modes at multiples of the ordering vector, $\pm nQ$.\cite{Sousa2008}
However, the electric dipole active isotropic exchange modulation mechanism that couples light to the spin wave at Brillouin zone boundary\cite{Aguilar2009} is apparently not active because optically active spin modes in \LCO/ are  below 30\wn/ while  the zone boundary energy of the spin waves in the $b$ direction is more than  60\wn/, Ref. \onlinecite{Masuda2005}.

\begin{table}[tb]
\caption{\label{LiTable}
Spin  excitations observed in the THz absorption spectra of \LCO/  in the order of increasing  energies at $T=3.5$\,K and  $B_0=0$\,T.
Best fit Lorentzian parameters -- center energy $\hbar\omega_0$ (cm$^{-1}$), linewidth at half maximum $\gamma$ (cm$^{-1}$) and line area  (cm$^{-2}$) are listed for three different orientations of $\mathbf{E}_1 $ and $\mathbf{B}_1 $.
}
\begin{tabular}{l|ddd|ddd|ddd}
\hline\hline 
 & \multicolumn{3}{c}{$\mathbf{E}_1 \!\parallel\! a(b)$} \vline& \multicolumn{3}{c}{$\mathbf{E}_1 \!\parallel\! \mathbf{c}$} \vline & \multicolumn{3}{c}{$\mathbf{E}_1 \!\parallel\! a(b)$} \\
 & \multicolumn{3}{c}{$\mathbf{B}_1 \!\parallel\! b(a)$} \vline& \multicolumn{3}{c}{$\mathbf{B}_1 \!\parallel\! a(b)$} \vline & \multicolumn{3}{c}{$\mathbf{B}_1 \!\parallel\! \mathbf{c}$} \\
\cline{2-10}
& \multicolumn{1}{c}{$\hbar\omega_0$} &  \multicolumn{1}{c}{$\gamma$} & \multicolumn{1}{c}{$A$}\vline & \multicolumn{1}{c}{$\hbar\omega_0$} &  \multicolumn{1}{c}{$\gamma$} & \multicolumn{1}{c}{$A$} \vline &  \multicolumn{1}{c}{$\hbar\omega_0$} &  \multicolumn{1}{c}{$\gamma$} & \multicolumn{1}{c}{$A$} \\
\hline
$ \mathrm{A}$ & 6.6 & 0.3 & 0.5 & 6.6 & 0.4 & 1.0 & 6.6 & 0.3 & 1.1 \\
$ \mathrm{B}$ & - & - & - & 8.3 & 0.2 & 0.1 & 8.3 & 0.2 & 0.1 \\
$ \mathrm{C}^{1}$ & 11.4 & 0.3 & 0.2 & 11.4 & 0.4 & 1.0 & 11.4 & 0.4 & 1.8 \\
$ \mathrm{C}^{2}$ & 12.1 & 0.5 & 1.5 & 12.0 & 0.5 & 1.7 & 12.0 & 0.5 & 1.3 \\
$ \mathrm{D}$ & 19.1 & 0.8 & 0.3 & 19.2 & 0.7 & 0.4 & 19.2 & 0.8 & 0.6 \\
$ \mathrm{E}^{1}$ & 23.8 & 0.3 & 0.1 & 23.9 & 0.5 & 0.7 & 23.9 & 0.5 & 2.0 \\
$ \mathrm{E}^{2}$ & 24.5 & 0.7 & 1.4 & 24.5 & 0.6 & 1.5 & 24.4 & 0.6 & 1.1 \\
$ \mathrm{F}$ & 28.6 & 1.2 & 0.2 & 28.4 & 1.2 & 0.3 & 28.4 & 0.7 & 0.6 \\
\hline\hline
\end{tabular}
\end{table}

The polarization dependence of THz absorption spectra is presented in Fig.\,\ref{LiAbsPolDep} and fit results of the absorption lines are given in Table\,\ref{LiTable}.
Our measurement rules out both $ab$ and $bc$ plane spin cycloids within the selection rules discussed above and outlined in Fig.\,\ref{fig1}b.
For the \LCO/ crystal structure a model where there are two cycloids corresponding to two zig-zag chains, tilted out from the $bc$ plane by an angle $\pm \theta$, seems natural.
The absorption line area for the $ \mathbf{B}_1\!\parallel\! \mathbf{c} $ polarization becomes $A_c=4A_0\cos^{2}\theta$ with $A_0$ standing for the line area when $ \mathbf{B}_1$ is in the cycloid plane.
The factor 4 is  two cycloids per unit cell  times two twin  domains.
For $ \mathbf{B}_1$ in the $ab$ plane the line area is $A_{ab}=2A_0( \sin^{2}\theta+1)$ where the first term in brackets is from  $ \mathbf{B}_1\!\perp\!\mathbf{b} $ domains and the second from $ \mathbf{B}_1\!\parallel\!\mathbf{b} $ domains. 
Lines C$ ^{2} $ and E$ ^{2}$ satisfy this model and give   $\theta=41\pm1^{\circ}$ for the cycloid tilt angle which is  close to crystallographic tilt angle $38.4^{\circ}$ of the zig-zag chains.
But the tilted cycloid model  does not explain the polarization dependence of other lines.
Moreover, the magnetic and the electric dipole approximations do not hold for lines A, C$ ^{1} $ and E$ ^{1}$.
They are weakest when light  propagates parallel to  the $c$ axis and  stronger when light propagates perpendicular to the $c$ axis.
In the dipole approximation the absorption line area depends on the orientation of light polarization with respect to the crystal axes and not on the orientation of the $\mathbf{k}$ vector of light.
A possible explanation of this $\mathbf{k}$ dependence  is the circular dichroism or electric quadrupolar transitions, but this needs further theoretical study.

Two cases of magnetic resonance in helical structures have been treated theoretically in the spin wave approximation, $\mathbf{B}_0$ perpendicular to the spin rotation plane and $\mathbf{B}_0$ in the plane of spin rotation.\cite{Cooper1962,Cooper1963,Nagamiya1967,Kataoka1987}
These results cannot be directly adapted to our experiment since the magnetic field would be perpendicular to the cycloidal plane only for an $ab$ or $bc$ plane circular cycloid in \LCO/. 
Given the  complexity and quantum nature of the problem, the theoretical description of spin wave modes in \LCO/ is not within the scope of the current paper. 
The number of modes is larger in Fig.\ref{BdepLinePos}b than in Fig.\,\ref{BdepLinePos}a because two crystal domains, $\mathbf{B}_0 \!\parallel\! \mathbf{b}$ and $\mathbf{B}_0 \!\perp\! \mathbf{b}$,  contribute to the absorption; in $\mathbf{B}_0\!\parallel\! \mathbf{c}$ twin domains are equal with respect to the field.
Some changes are evident in the THz absorption spectra around 2\,T, Fig.\,\ref{BdepLinePos}b.
This is due to  reordering of  spins  when $\mathbf P$  flops from $c$ to $a$ axis.
Line A splits between 1 and 2\,T and lines C$ ^{1} $ and C$ ^{2} $ acquire dispersing modes C$ ^{1}_\parallel$ and C$ ^{2}_\parallel$, Fig.\,\ref{BdepLinePos}c.
We assign the C$ ^{1}_\parallel $ and C$ ^{2}_\parallel$ modes together with new modes A$ _{\parallel} $ and D$ _{\parallel} $ to domains where $\mathbf  B_0 \!\parallel\! \mathbf  b$ since this is the field direction where $\mathbf  P$ is flopped.
There is some evidence,  Fig.\,\ref{BdepLinePos}c, that A$ _{\!\parallel\!} $ and G modes anti-cross and therefore the G mode may belong to the $\mathbf  B_0 \!\parallel\! \mathbf  b$  domains.

Electromagnons contribute to the zero frequency dielectric constant as
$\Delta\varepsilon=\varepsilon(0) - \varepsilon_\infty = \sum_j (\omega_p^j / \omega_0^j)^2$,
where the sum runs over all spin wave modes observed below 35\wn/ in the $\mathbf{E}_1\!\parallel\! \mathbf{a}$ THz absorption spectrum, $\omega_0^j $ and $ \omega_p^j $ are the $j$-th mode resonance energy and plasma frequency, both in wavenumber units ($ \omega_p^j $ is related to line area by $\pi\omega_p^j = \sqrt{A n}$ with   refraction index $n=3.36$ for $\mathbf{E}_1\!\parallel\! \mathbf{a}$);
$ \varepsilon_\infty $ contains the contribution from all other modes above 35\wn/ and is taken $T$-independent below 30\,K.\cite{Papagno2006}
Assuming that all the oscillator strength is due to electric dipole activity we can set an upper limit to $\Delta\varepsilon_a = 0.024$ using $\mathbf{E}_1\!\parallel\! \mathbf{a(b)}, \mathbf{B}_1\!\parallel\! \mathbf{c}$ data from Table\,\ref{LiTable}.
A measurement at 28\,kHz gave $\Delta\varepsilon_a=0.1$, Ref.\,\onlinecite{Park2007}, which is four times larger.
However, the $B_0$ dependence shows (Fig.\,\ref{BdepLinePos}c) that there is an additional mode (H) which shifts into our measurement window at linear rate $0.522\pm0.005$\,cm$^{-1}$T$^{-1}$ with a zero-field  intercept at $1.03\pm0.05$\wn/. 
Assuming the same oscillator strength for the zero-field resonance as is above 5\,T we get that the contribution of this 1\wn/ mode to the dielectric constant is $\Delta\varepsilon_a = 0.46$.
The actual number might be smaller because it is not known if, first, $\omega_p=\mathrm{const}$, and second, whether $\omega_0\propto B_0$ persists down to zero field.
A mode at 1\wn/, when extrapolated to 0\,T, has been detected with an electron spin resonance technique although with nonlinear $B_0$ dependence.\cite{Vorotynov1998}


In conclusion, the analysis of the  THz absorption spectra within the linear spin wave theory shows that the spin order is not  $ab$ nor $bc$ plane cycloid. 
The cycloid tilt angle $41\pm 1^{\circ}$ out of the $ bc$ plane is inferred from the polarization dependence of two modes assuming they are magnetic dipole active.
For other modes the polarization dependence is not compatible with magnetic or electric dipole activity of a cycloidal spin chain.
We have evidence that there is a spin wave mode at 1\wn/ with sufficient oscillator strength to be an electromagnon.
Understanding the origin of the large number of optically active  spin excitations and their interaction with magnetic field in a quantum spin system such as \LCO/ requires further theoretical investigation.

We thank M.\,Mostovoy for useful discussions.
Support by EstSF grants  5553, 6138, and 7011 is acknowledged. 
Work at Rutgers was supported by the DE-FG02-07ER46382.

\bibliographystyle{apsrev}


\end{document}